\documentclass[aps,prd,reprint,nofootinbib,showkeys,showpacs,superscriptaddress,longbibliography,floatfix]{revtex4-1}

\pdfoutput=1

\usepackage{amsmath}
\usepackage{amssymb}
\usepackage[american]{babel}
\usepackage{booktabs}
\usepackage{dcolumn}  
\usepackage[T1]{fontenc}  
\usepackage{graphicx}  
\usepackage[colorlinks]{hyperref}  
\usepackage[utf8]{inputenc}  
\usepackage{txfonts}  
\usepackage{url}
\usepackage[dvipsnames]{xcolor}
\usepackage{float}

\usepackage{tikz}

\usetikzlibrary{arrows,shapes,positioning,shadows,trees}

\tikzset{
	basic/.style  = {draw, text width=2cm, drop shadow, font=\sffamily, rectangle},
	root/.style   = {basic, rounded corners=2pt, thin, align=center,
		fill=green!30},
	level 2/.style = {basic, rounded corners=6pt, thin,align=center, fill=green!60,
		text width=8em},
	level 3/.style = {basic, thin, align=left, fill=pink!60, text width=6.5em}
}

\usetikzlibrary{arrows,decorations.pathmorphing,backgrounds,fit,positioning,shapes.symbols,chains}

\newcolumntype{d}[1]{D{.}{.}{#1}}
\newcolumntype{v}[1]{D{,}{,\ }{#1}}

\makeatletter

\newcommand{\Rmnum}[1]{\expandafter\@slowromancap\romannumeral #1@}
\makeatother

\hypersetup{
	citecolor = red,
	linkcolor = blue,
	urlcolor = magenta
}

\begin{document}

\title{Warm $\beta$-exponential inflation and the swampland conjectures}

\author{F. B. M. dos Santos}
\email{felipe.santos.091@ufrn.edu.br}
\affiliation{Universidade Federal do Rio Grande do Norte,
	Departamento de F\'{\i}sica, Natal - RN, 59072-970, Brasil}

\author{R. Silva}
\email{raimundosilva@fisica.ufrn.br}
\affiliation{Universidade Federal do Rio Grande do Norte,
	Departamento de F\'{\i}sica, Natal - RN, 59072-970, Brasil}
\affiliation{Departamento de F\'{\i}sica, Universidade do Estado do Rio Grande do Norte, Mossor\'o, 59610-210, Brasil}

\author{S. Santos da Costa}
\email{simony.santosdacosta@pi.infn.it}
\affiliation{Istituto Nazionale di Fisica Nucleare (INFN) Sezione di Pisa, Largo B. Pontecorvo 3, 56127 Pisa, Italy}

\author{M. Benetti}
\email{micol.benetti@unina.it}
\affiliation{Scuola Superiore Meridionale, Largo San Marcellino 10, 80138 Napoli, Italy}
\affiliation{Istituto Nazionale di Fisica Nucleare (INFN) Sezione di Napoli, Complesso Universitario di Monte Sant'Angelo,  Edificio G, Via Cinthia, I-80126, Napoli, Italy}

\author{J. S. Alcaniz}
\email{alcaniz@on.br}
\affiliation{Departamento de Astronomia, Observatório Nacional, Rua General José Cristino 20921-400, Rio de Janeiro-RJ, Brasil}

\begin{abstract}
     	We investigate theoretical and observational aspects of a warm inflation scenario driven by the $\beta$-exponential potential, which generalizes the well-known power law inflation. In such a scenario, the decay of the inflaton field into radiation happens during the inflationary phase. In our study, we consider a dissipation coefficient ($\Gamma$) with cubic dependence on the temperature ($T$) and investigate the consequences in the inflationary dynamics, focusing on the impact on the spectral index $n_s$, its running $n_{run}$ and tensor-to-scalar ratio $r$. We find it possible to realize inflation in agreement with current cosmic microwave background data in weak and strong dissipation regimes. We also investigate theoretical aspects of the model in light of the swampland conjectures, as warm inflation in the strong dissipation regime has been known as a way to satisfy the three conditions currently discussed in the literature. We find that when $\Gamma\propto T^3$, the $\beta$-exponential model can be accommodated into the conjectures.
\end{abstract}

\maketitle

\section{Introduction}

The $\Lambda$CDM model, combined with the idea of primordial inflation, constitutes a remarkable description of the universe evolution from very early to late times. In particular, inflation solves some of the problems that arise in the big bang theory by assuming a rapid expansion of the universe while generating initial conditions for the subsequent cosmic evolution \cite{Starobinsky:1980te,Sato:1980yn,Guth1981,Linde:1981mu,Linde:1983gd,Sato:2015dga}. In the inflationary framework, the \textit{inflaton} field is responsible for the early accelerated expansion, whose evolution is driven by a specific potential function. Naturally, over the years, many possible candidates appeared \cite{Martin:2013tda}, from which some of them are viewed as viable models, as they agree with current observations provided by cosmic microwave background (CMB) experiments \cite{Aghanim:2018eyx,Planck:2018jri}. 

After inflation, however, one needs to direct the attention to a \textit{reheating} period, that connects the inflationary era to radiation dominance \cite{Abbott:1982hn,Albrecht:1982mp}. In this epoch, the inflaton couples to other fields such that the remaining energy is converted to create new particles that compose the radiation energy density. While much progress has been made in the description of this era and its connection with CMB data \cite{Dolgov:1989us,Traschen:1990sw,Greene:1997fu,Kofman:1994rk,Kofman:1997yn,Greene:1997ge,Dufaux:2006ee,Abolhasani:2009nb,Munoz:2014eqa,Dai:2014jja,Cook:2015vqa,Eshaghi:2016kne,Drewes:2017fmn,Saha:2020bis}, the exact mechanism is still unknown since many factors may appear, and they can be very dependent on the inflationary model in consideration. In this scenario of \textit{cold inflation}, the coupling to other fields is neglected until inflation ends. On the other hand, an alternative is to consider that this coupling is relevant during inflation, which characterizes the \textit{warm inflation} picture \cite{Berera:1995wh,Berera:1995ie,Yokoyama:1998ju}. The coupling of the inflaton to other fields creates a thermal bath in which the production of relativistic particles reheats the universe so that the universe can go smoothly to a radiation-dominated era by the end of inflation. Indeed, a dissipative term in the equations of motion provides extra friction, which implies a modification in the description of the accelerated expansion and, consequently, in the observational predictions. 

The warm inflation picture has been widely studied in recent literature. In general, from a phenomenological perspective, models driven by potentials that are disfavored by data in the cold inflation picture may become viable as, for example, scenarios described by monomial potentials \cite{Bartrum:2013fia,Arya:2017zlb,Bastero-Gil:2017wwl,Arya:2018sgw,Bastero-Gil:2021fac,Rosa:2019jci} (See \cite{Benetti:2016jhf,Reyimuaji:2020bkm} for the predictions of other known models). This happens because as the dissipation coefficient introduces another term of friction in the equation of motion of the inflaton, an extra factor appears in the slow-roll parameters so that they can be suppressed more effectively, even when the potential is steep \cite{Das:2020xmh,Motaharfar:2021egj}. From a more fundamental point of view, warm inflation might arise from concrete particle physics scenarios \cite{Bastero-Gil:2016qru,Bastero-Gil:2018yen,Berghaus:2019whh,Levy:2020zfo}, being able to sustain particle production leading to a `graceful exit' to the radiation era.

This work investigates the warm inflation scenario driven by a class of  $\beta$-exponential potentials that generalizes the well-known power law inflation \cite{Alcaniz:2006nu}. As shown in \cite{Santos:2017alg}, such a model can arise from brane dynamics and showed a good agreement with Planck 2015 data. An updated analysis with Planck 2018 and clustering data showed that the model with a non-minimal coupling of the field with gravity seems to be a more viable approach for this class of models \cite{dosSantos:2021vis}. Here we investigate how the predictions of the $\beta$-exponential inflationary model change when considering the warm inflation picture, as recent studies have investigated the viability of exponential potentials in this context. For example, in \cite{Das:2019acf}, a pure exponential potential was considered in the strong dissipation regime, with a dissipative coefficient $\Gamma\propto T^3$; a coupling of the type was motivated in \cite{Berghaus:2019whh}, where the assumption is that the scalar field has an axionic coupling to gauge bosons. The application to an exponential potential, as investigated in \cite{Das:2019acf}, showed that either inflation still would not end by violation of the slow-roll conditions or the predicted spectral index was too red-tilted in the strong dissipative regime. The distortion in the exponential form caused by the $\beta$-exponential function may address both issues. It is worth noting that another generalization of the exponential function was considered in \cite{Das:2020xmh,Lima:2019yyv}, showing that runaway-type potentials are also an option in the warm inflation picture. In particular, the tensor-to-scalar ratio becomes significantly suppressed if one wants to achieve the central Planck value for $n_s$. 

Another point of investigation concerns the recently proposed \textit{swampland conjectures} \cite{Kachru:2003aw,Obied:2018sgi,Ooguri:2006in,Palti:2019pca}. In this concern, some works constrain scalar field theories based on the assumption that they can be embedded in more general theories, such as string theory \cite{Agrawal:2018own, Garg:2018reu,Achucarro:2018vey,Motaharfar:2018zyb, Kamali:2019hgv, Benetti:2019smr}. A discussion started from the difficulty in obtaining de Sitter vacua in these theories \cite{Danielsson:2018ztv,Dine:2020vmr,Kachru:2003aw}; therefore, in order for a model to be theoretically consistent, it should obey certain limits to stay in the \textit{landscape} of well-motivated scenarios. In particular, it was shown in \cite{Das:2018hqy,Motaharfar:2018zyb,Das:2019acf,Kamali:2019xnt,Berera:2019zdd,Das:2019hto,Das:2020xmh,Brandenberger:2020oav} that warm inflation realized in the strong dissipative regime makes it possible for all current conjectures to be satisfied. The warm inflation idea combined with extensions of the canonical picture as done in Refs. \cite{Benetti:2019kgw,Kamali:2019xnt,Mohammadi:2020vgs} can also be considered to recover concordance with observations. Adding to these recent interesting studies, in this work we want to determine how far from a simple exponential form one may go to provide reasonable predictions. This way, we study if the $\beta$-exponential model can be another option in a warm inflation construction in which the strong regime can be realized while being consistent with CMB data and its impact on the swampland conjectures.

This work is organized in the following manner: in section \ref{sec2}, we review warm inflation and the respective slow-roll equations. In Section \ref{sec3}, we introduce the $\beta$-exponential model into the warm inflation framework, while in Section \ref{sec4}, we discuss if the model is consistent with the swampland conjectures. To conclude, in Section \ref{sec5}, we present our considerations.

\section{Warm inflation}\label{sec2}

The dissipation of inflaton into other particles is often modeled by the presence of a dissipation coefficient $\Gamma$ in the equation of motion of the scalar field. It means that the Klein-Gordon equation for $\phi$ becomes
\begin{gather}
	\ddot\phi + \left(3H+\Gamma\right)\dot\phi + V_{,\phi}=0,
	\label{1}
\end{gather}
during inflation. Here, a dot denotes a derivative in time, while the subscript $_{,\phi}$ represents a derivative w.r.t. the field. We see that the additional term with $\Gamma$ constitutes an additional source of friction added to the Hubble one. Since the field decays into radiation, the energy density evolution comes from the conservation of the energy-momentum tensor as
\begin{gather}
	\dot\rho_r + 4H\rho_r = \Gamma\dot\phi^2, \quad \dot\rho_\phi + 3H(\rho_\phi + P_\phi) = -\Gamma\dot\phi^2,
	\label{2}
\end{gather}
where we note that the term proportional to $\Gamma$ represents the energy transferred to the radiation particles from the inflaton. To close the set of equations, we need the Friedmann equation, which gives us the background expansion
\begin{gather}
	3H^2M_{p}^2 = \frac{1}{2}\dot\phi^2 + V + \rho_r,
	\label{3}
\end{gather}
with $M_{p}=\frac{1}{\sqrt{8\pi G}}$ being the reduced Planck mass. The usual procedure in single scalar field inflation is to apply the slow-roll approximation, in which the field slowly rolls down the potential; this is achieved by neglecting higher-order derivatives in the equations of motion and assuming that the potential dominates the energy budget of the field. As a consequence, Eqs. (\ref{1}-\ref{3}) reduce to
\begin{gather}
	\dot\phi \simeq -\frac{V_{,\phi}}{3H(1+Q)},\quad H^2 \simeq \frac{V}{3M_p^2},\nonumber\\
	4H\rho_r\simeq \Gamma\dot\phi^2.
	\label{4}
\end{gather}
Here, we have introduced the ratio $Q\equiv\frac{\Gamma}{3H}$ as standard practice, and we have neglected $\dot\rho_r$ by assuming that the thermal equilibrium of the bath is quickly achieved. We note that the value of $Q$ determines how effective the dissipation is: For $\Gamma<H$, we have $Q<1$, characterizing the \textit{weak} dissipative regime; on the other hand, if $\Gamma>H$, then $Q>1$, and inflation proceeds in the \textit{strong} dissipative regime. In the same manner as the cold inflation picture, one can derive slow-roll parameters expressed as
\begin{gather}
	\begin{split}
    \epsilon_W & \equiv \frac{\epsilon_V}{1+Q} = \frac{M_p^2}{2(1+Q)}\left(\frac{V_{,\phi}}{V}\right)^2, \\
    \eta_W &\equiv \frac{\eta_V}{1+Q}=\frac{M_p^2}{1+Q}\left(\frac{V_{,\phi\phi}}{V}\right), \\
    \beta_W &\equiv \frac{M_p^2}{1+Q}\left(\frac{\Gamma_{,\phi}V_{,\phi}}{\Gamma V}\right),
    \label{5}  
    \end{split}  
\end{gather}
and during inflation, $\epsilon_W,|\eta_W|,|\beta_W|\ll 1$. One interesting aspect of slow-roll parameters in Eq. (\ref{5}) is that the slow-roll regime can be properly achieved even for steep potentials. When $Q$ is relevant, all three parameters can take smaller values, thus making the slow-roll regime possible. The dissipation coefficient $\Gamma$ is usually dependent on the temperature of the bath; therefore, let us determine it, especially since it is connected to the final temperature that starts the radiation era. By assuming quick thermalization, the radiation energy density can be written in terms of its temperature as
\begin{gather}
	\rho_r = \tilde{g}_\star T^4,
	\label{6}
\end{gather} 
where $\tilde{g}_\star\equiv\frac{\pi^2g_\star}{30}$ and $g_\star$ is the number of relativistic degrees of freedom of the fields during inflation. This thermal equilibrium implies that $T>H$, along with the slow-roll conditions, is necessary for warm inflation. We can combine the expression for $\rho_r$ in Eq. (\ref{4}) with the one in Eq. (\ref{6}), to obtain the ratio $T/H$ as a function of $Q$ and $\phi$
\begin{gather}
    \frac{T}{H} = \left[\frac{9QM_p^6V_{,\phi}^2}{4\tilde g_\star(1+Q)^2V^3}\right]^{1/4},
    \label{7}
\end{gather}
where we have also used the equations for $\dot\phi$ and $H$ in Eq. (\ref{4}).

The scalar power spectrum is also affected by dissipation during inflation. It has the form \cite{Ramos:2013nsa,Bartrum:2013fia}
\begin{gather}
	\Delta^2_\mathcal{R} = \left(\frac{H_\star^2}{2\pi\dot\phi_\star}\right)^2\left( 1 + 2n_{BE} + \frac{2\sqrt{3}\pi Q_\star}{\sqrt{3 + 4\pi Q_\star}}\frac{T_\star}{H_\star} \right)G(Q_\star),
	\label{8}
\end{gather}
where $n_{BE}=\frac{1}{e^{H_\star/T_\star}-1}$ is the Bose-Einstein distribution function, and $G(Q_\star)$ is an enhancement term that has been argued to be present depending on the dependence of $\Gamma$ on the temperature \cite{Bastero-Gil:2011rva,Bastero-Gil:2016qru,Ramos:2013nsa}, and it arises from the interaction of the inflaton with radiation. In this work, we consider a cubic dependence on $T$; a numerical fit of $G(Q)$ for this case has been found as \cite{Bastero-Gil:2011rva,Bastero-Gil:2016qru}
\begin{gather}
	G_{cubic}(Q_\star) = 1 + 4.981 Q_\star^{1.946} + 0.127 Q_\star^{4.330}.
	\label{9}
\end{gather}
We note that by taking the limits $T\rightarrow0$, $Q\rightarrow0$ in Eqs. (\ref{8}) and  (\ref{9}), one achieves the cold inflation limit. All quantities are computed at the pivot scale $k=k_\star$, at which the CMB scale leaves the horizon, for which the amplitude of the scalar power spectrum is estimated as $\log(10^{10}\Delta_\mathcal{R}^2)=3.044 \pm 0.014$ \cite{Aghanim:2018eyx}. As for the tensor power spectrum, it is argued that we can approximate it as having the same form as in cold inflation \cite{Ramos:2013nsa,Benetti:2016jhf}
\begin{gather}
    \Delta^2_h = \frac{2H^2}{\pi^2M_p^2},
    \label{10}
\end{gather}
so that we can readily write the tensor-to-scalar ratio as
\begin{gather}
	r = \frac{\Delta^2_h}{\Delta^2_\mathcal{R}}.
	\label{11}
\end{gather}
The spectral index $n_s$ and its running $n_{run}$ can be derived from (\ref{8}) as
\begin{gather}
    n_s-1=\frac{d\operatorname{log}\Delta^2_\mathcal{R}}{d\operatorname{log}k}\simeq \frac{d\operatorname{log}\Delta^2_\mathcal{R}}{dN}, \quad n_{run}=\frac{d^2\operatorname{log}\Delta^2_\mathcal{R}}{d\operatorname{log}k^2}\simeq \frac{d^2\operatorname{log}\Delta^2_\mathcal{R}}{dN^2},
    \label{12}
\end{gather}
where $N$ is the number of e-folds. In Appendices \ref{appA} and \ref{appB} we show a general manner for deriving an expression for $n_s$ and $n_{run}$ for a given dissipation coefficient $\Gamma$. 

\section{Warm $\beta$-inflationary model}\label{sec3}

\begin{figure*}
	\centering
	\includegraphics[width=7.5cm]{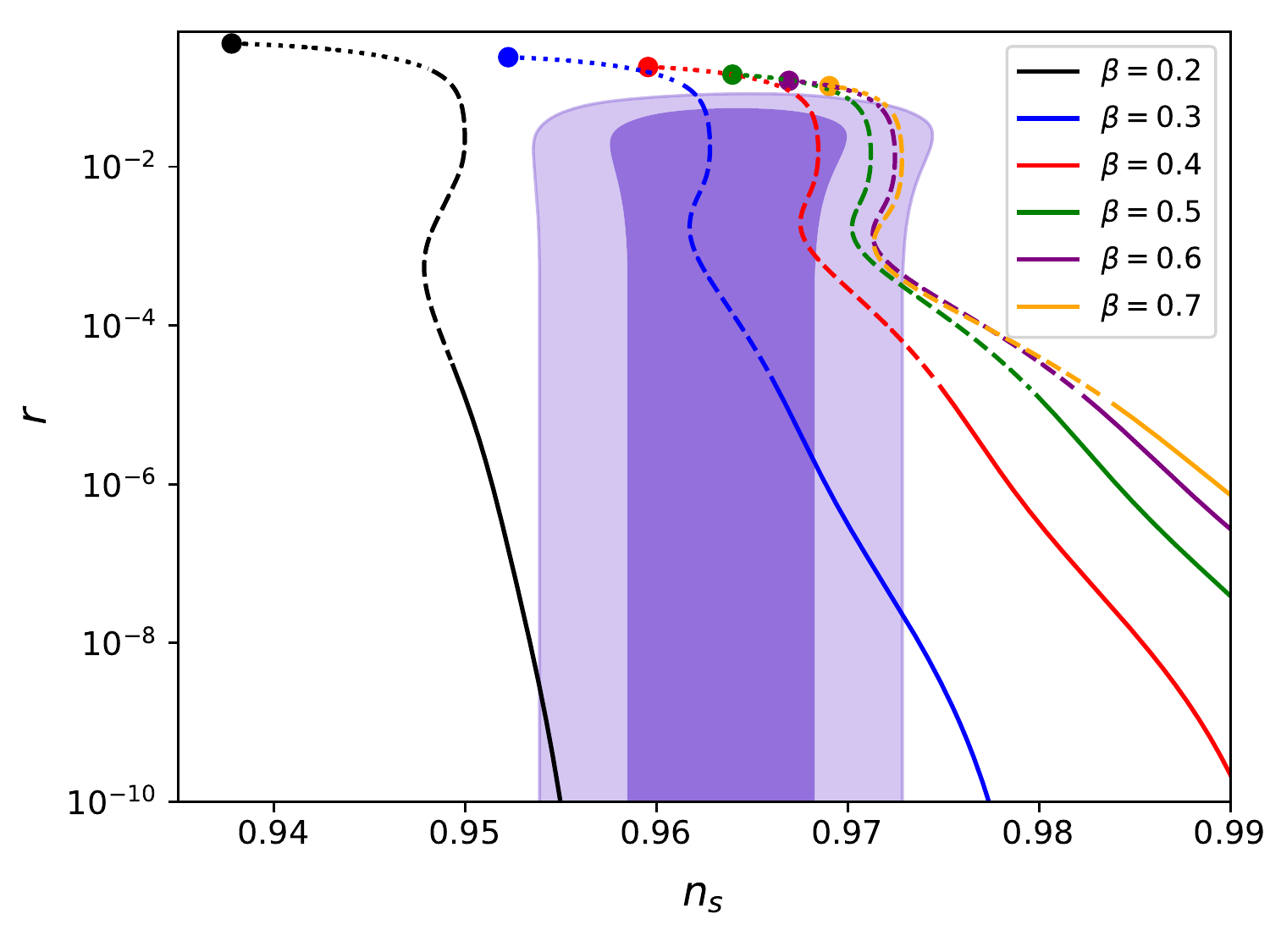}	
	\includegraphics[width=7.5cm]{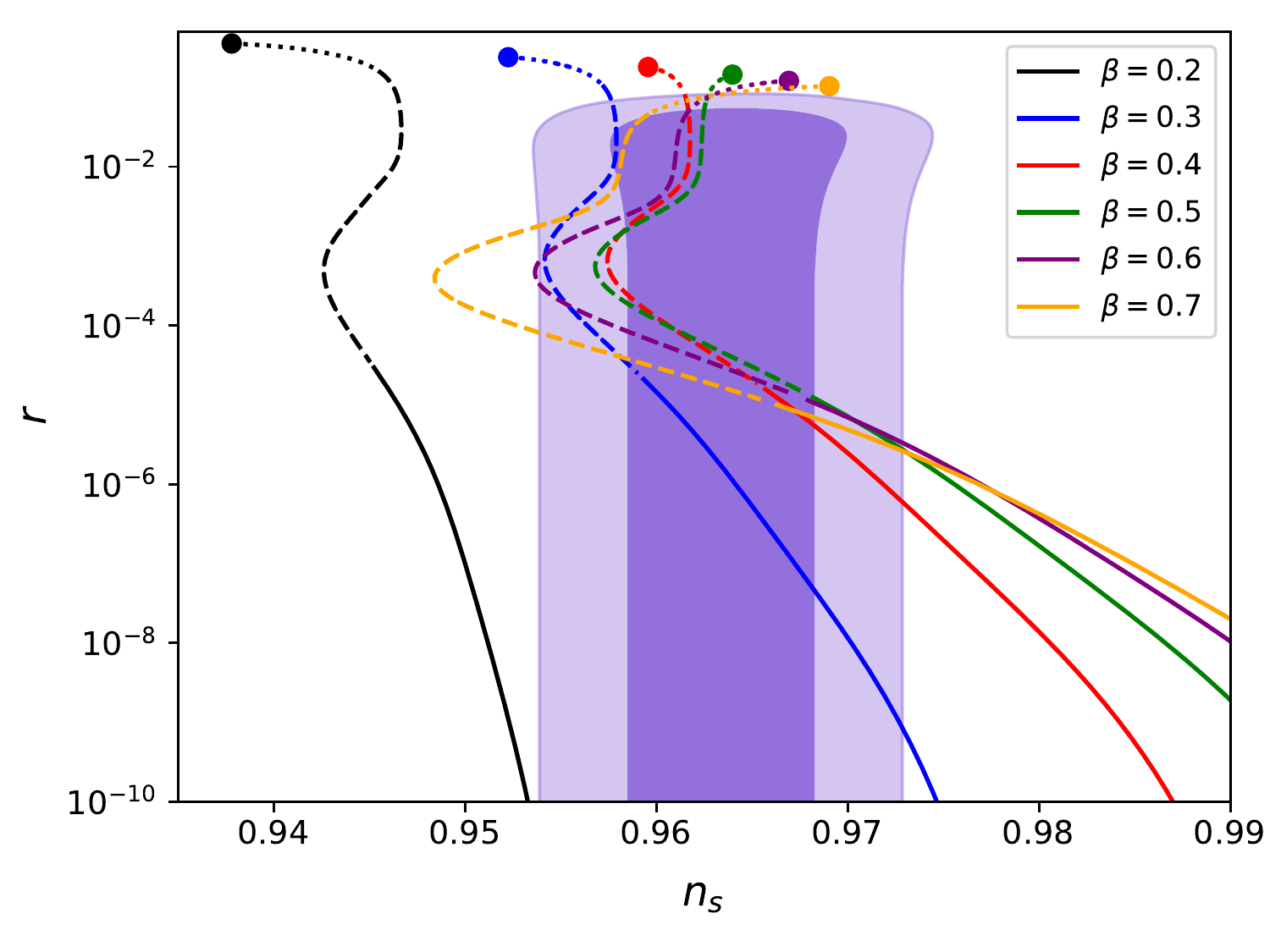}	
	\includegraphics[width=7.5cm]{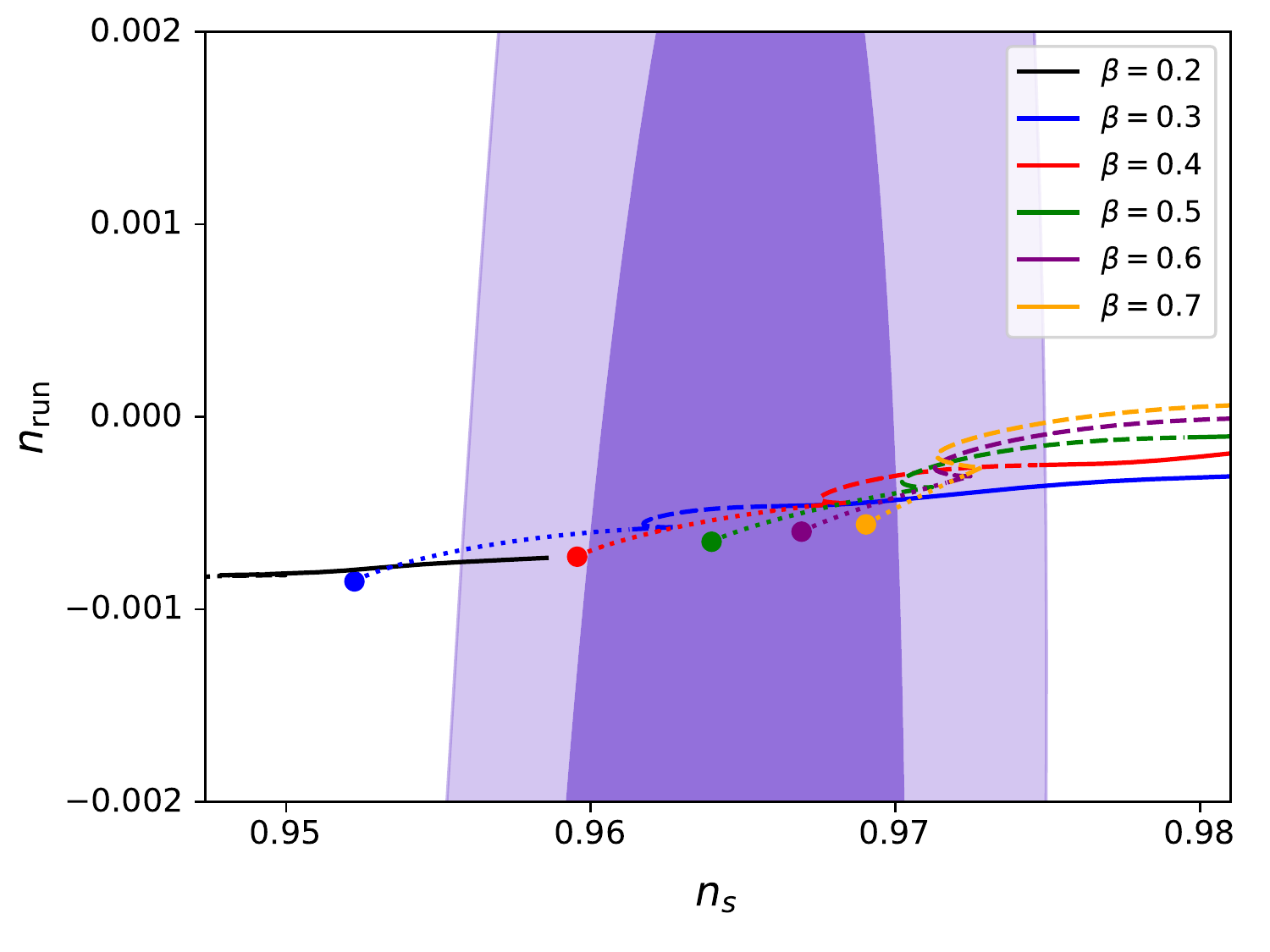}	
	\includegraphics[width=7.5cm]{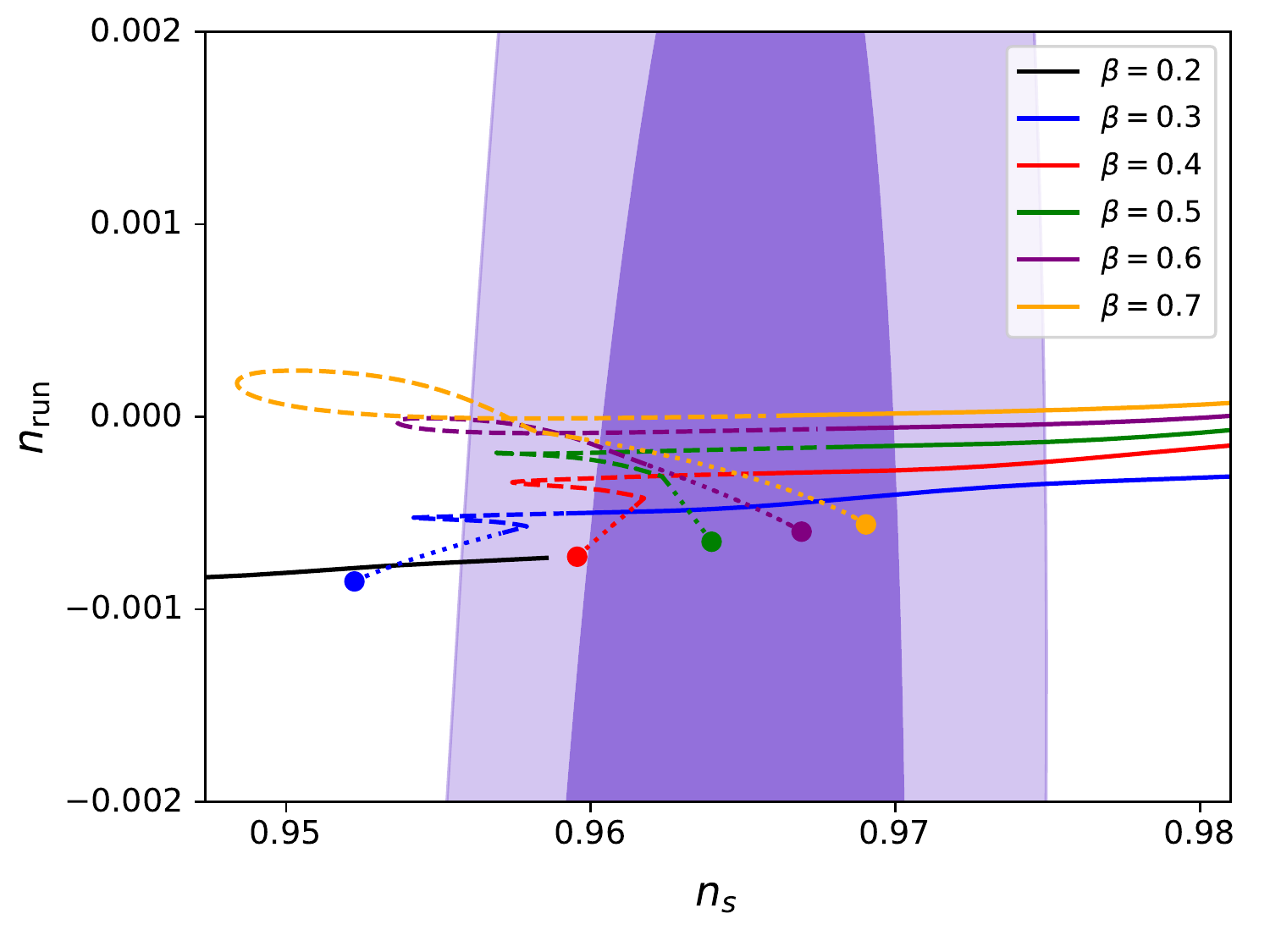}	
	\caption{Predictions of $n_s-r$ and $n_s-n_{run}$ planes for the $\beta$-exponential model with $\Gamma=C_\phi T^3/\phi^2$. We have chosen fixed values of $\beta$ while considering $\lambda=0.05$ (left panels) and $\lambda=0.07$ (right panels), with the curves varying with $Q_\star$. The dashed lines represent the weak dissipation regime, while the solid lines represent the strong regime; the dotted lines correspond to a regime where $T/H<1$. We have set $N_\star=55$.}
	\label{fig1}
\end{figure*}

The application we consider in this work is to the $\beta$-exponential model \cite{Alcaniz:2006nu}
\begin{gather}
	V(\phi)=V_0\left[1-\lambda\beta\frac{\phi}{M_p}\right]^{1/\beta},
	\label{13}
\end{gather}
where $\lambda$ is a dimensionless constant, and $\beta$ is another constant that controls the deviation from the pure exponential function, achieved for $\beta\rightarrow0$. We note that $\beta$ should not be confused with $\beta_W$, which is one of the slow-roll parameters in Eq. (\ref{5}). This model was first proposed as a phenomenological generalization of the exponential potential $V=V_0e^{-\lambda\frac{\phi}{M_p}}$, for which dissipative effects were studied in \cite{Yokoyama:1987an}.

The $\beta$-exponential potential is able to achieve the breakdown of the slow-roll regime, with the end of inflation by assuming $\epsilon_V=1$, and is able to predict the low values for the tensor-to-scalar ratio, r, observed by recent experiments increasing the $\beta$ value \cite{Alcaniz:2006nu}. Further investigations of the model resulted in a concordance at the $2\sigma$ level with the Planck 2015 $n_s-r$ data and reasonably favored results when the model was statistically compared with the $\Lambda$CDM one \cite{Santos:2017alg}. Also, a more fundamental theoretical motivation for the model was found from brane dynamics; as a result, the ratio $\beta/\lambda$ becomes associated with the brane tension $\sigma$. This relation imposes a constraint on both $\beta$ and $\lambda$, where essentially, $\beta$ must be larger than $\lambda$, with $\beta\geq1/2$. In \cite{Santos:2017alg,dosSantos:2021vis}, this limit is satisfied in the priors and numerical analysis results. Our analysis will also check if this constraint is still respected when computing the inflationary observables.

The slow-roll parameters for the model are
\begin{gather}
	\begin{split}
	\epsilon_w &= \frac{M_p^2\lambda^2}{2(1+Q)\left(1-\lambda\beta\frac{\phi}{M_p}\right)^2},\\
	\eta_w &= \frac{M_p^2\lambda^2(1-\beta)}{(1+Q)\left(1-\lambda\beta\frac{\phi}{M_p}\right)^2},\\
	\beta_w &= \epsilon_w\left(1 + 2\frac{\left(1-\lambda\beta\frac{\phi}{M_p}\right)}{\lambda}\frac{Q_{,\phi}}{Q}\right),
	\label{14}
	\end{split}	
\end{gather}
We determine the end of inflation for both $\phi$ and $Q$. The ratio $\frac{T}{H}$ is computed from Eq. (\ref{7}) as 
\begin{gather}
	\frac{T}{H} = \left[\frac{9}{4\tilde g_\star}\frac{Q}{(1+Q)^2}\left(\frac{M_p^4}{V_0}\right)\frac{\lambda^2}{\left(1-\lambda\beta\frac{\phi}{M_p}\right)^{1/\beta+2}}\right]^{1/4}
	\label{15}
\end{gather} 
where the dependence on $\Gamma$ is implicit on $Q$. Hereafter, we shall apply the model to a dissipation coefficient with cubic dependence on the temperature. The general strategy is as follows: we first isolate an expression for $Q_{end}$ from the slow-roll condition by setting $\epsilon_W=1$. Then, by finding the general relation between $Q$ and $\phi$ for each $\Gamma$, it is possible to compute a value for $\phi_{end}$ numerically for each set of parameters. Next, we use $Q_{end}, \phi_{end}$ as initial conditions for the first-order differential equations that are evolved back to $N_\star$, so that we obtain $Q_\star$ and $\phi_\star$ and finally compute $n_s$, $n_{run}$ and $r$ for a given $\lambda,\beta$ (Appendix \ref{appA}). Then we will see the impact of considering different $\lambda$ and $\beta$ in the $n_s-r$ and $n_s-n_{run}$ plane and the temperature at the end of inflation. Also, we will see whether inflation with this model is favored in the weak or strong regime to find if the model addresses the swampland conjectures to be discussed in Section \ref{sec4}.

The dissipation coefficient we will consider has a cubic dependence on the temperature, with the form
\begin{gather}
	\Gamma=C_\phi \frac{T^3}{\phi^2}.
	\label{16}
\end{gather}

This form can be motivated from a supersymmetric setting, where the correspondent superpotential involves interacting $\Phi$, $X$, and $Y$ superfields \cite{Hall:2004zr,Moss:2006gt,Bastero-Gil:2009sdq,Bastero-Gil:2010dgy,Bastero-Gil:2012akf}. It comes from the possibility of a two-stage decay $\phi\rightarrow\chi\rightarrow yy$, in which the field $\chi$ acts as an intermediate to decay into light particles $y$. The potential obtained from the general superpotential generates a mass for the bosonic and fermionic components of the field $X$, denoted by $\chi$, from which it is possible to derive a low-temperature approximation ($T \ll m_\chi$) for $\Gamma$, as studied in \cite{Moss:2006gt}. For large field multiplicities, and by knowing that $m_\chi^2=2g^2\phi^2$, one arrives at the form given by Eq. (\ref{16}) with $C_\phi\simeq \frac{1}{16\pi}h^2 N_Y N_X$ \cite{Bastero-Gil:2012akf}.    

Due to the coupling of the inflaton with radiation, one would expect corrections to the inflationary potential to appear, altering its form and potentially spoiling the slow-roll regime. For the setting described, it is possible to estimate both fermionic and bosonic contributions to the inflationary potential due to these couplings. In \cite{Hall:2004zr}, this issue was discussed, showing that while these corrections might be relevant, the effect on the slope of the potential is properly suppressed at a low-temperature regime. Thus, inflationary dynamics under (\ref{16}) can be well approximated as unaffected by radiative and thermal effects \cite{Bartrum:2013fia}.

\begin{figure*}
	\centering
	\includegraphics[width=7cm]{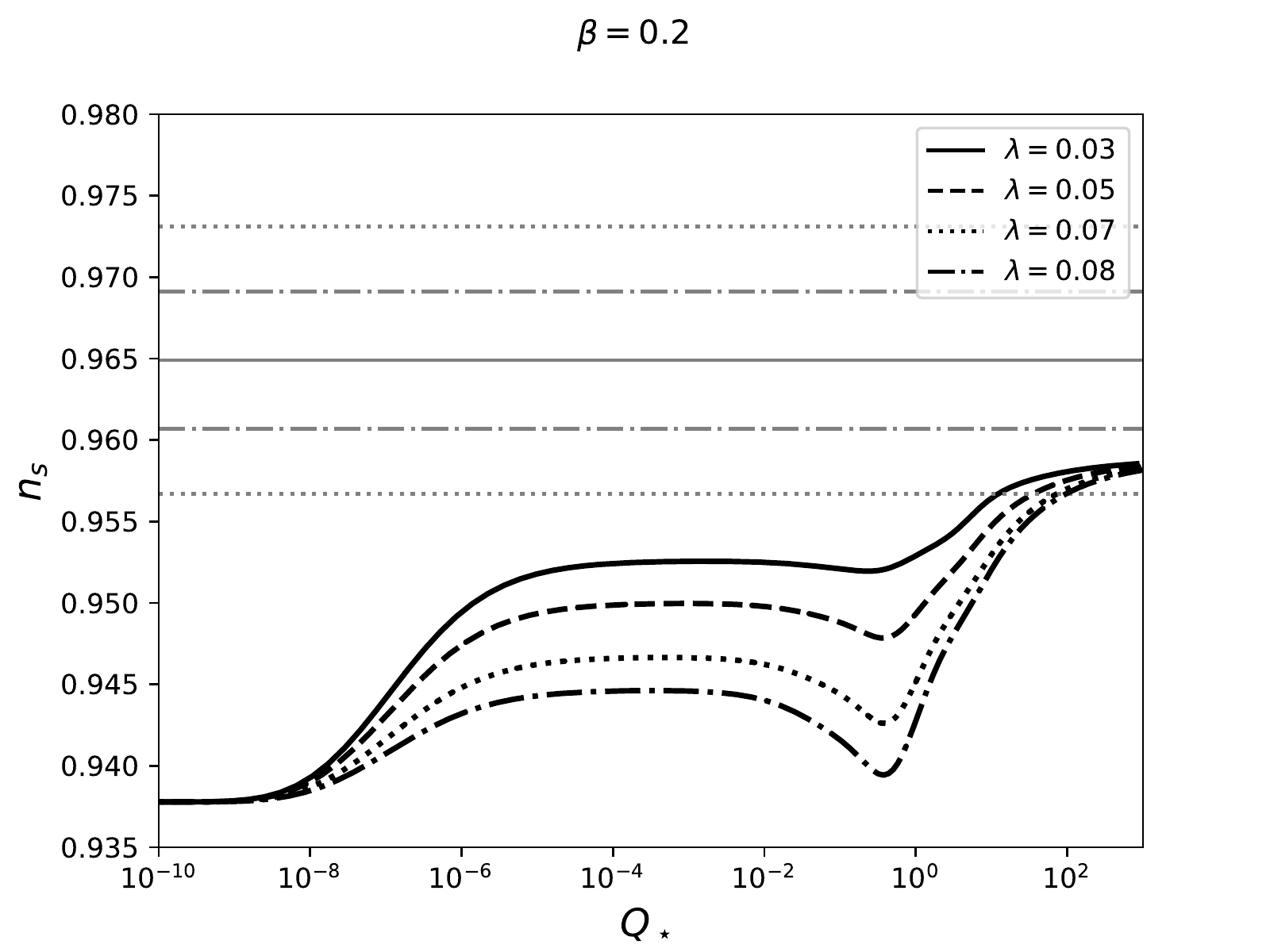}
	\includegraphics[width=7cm]{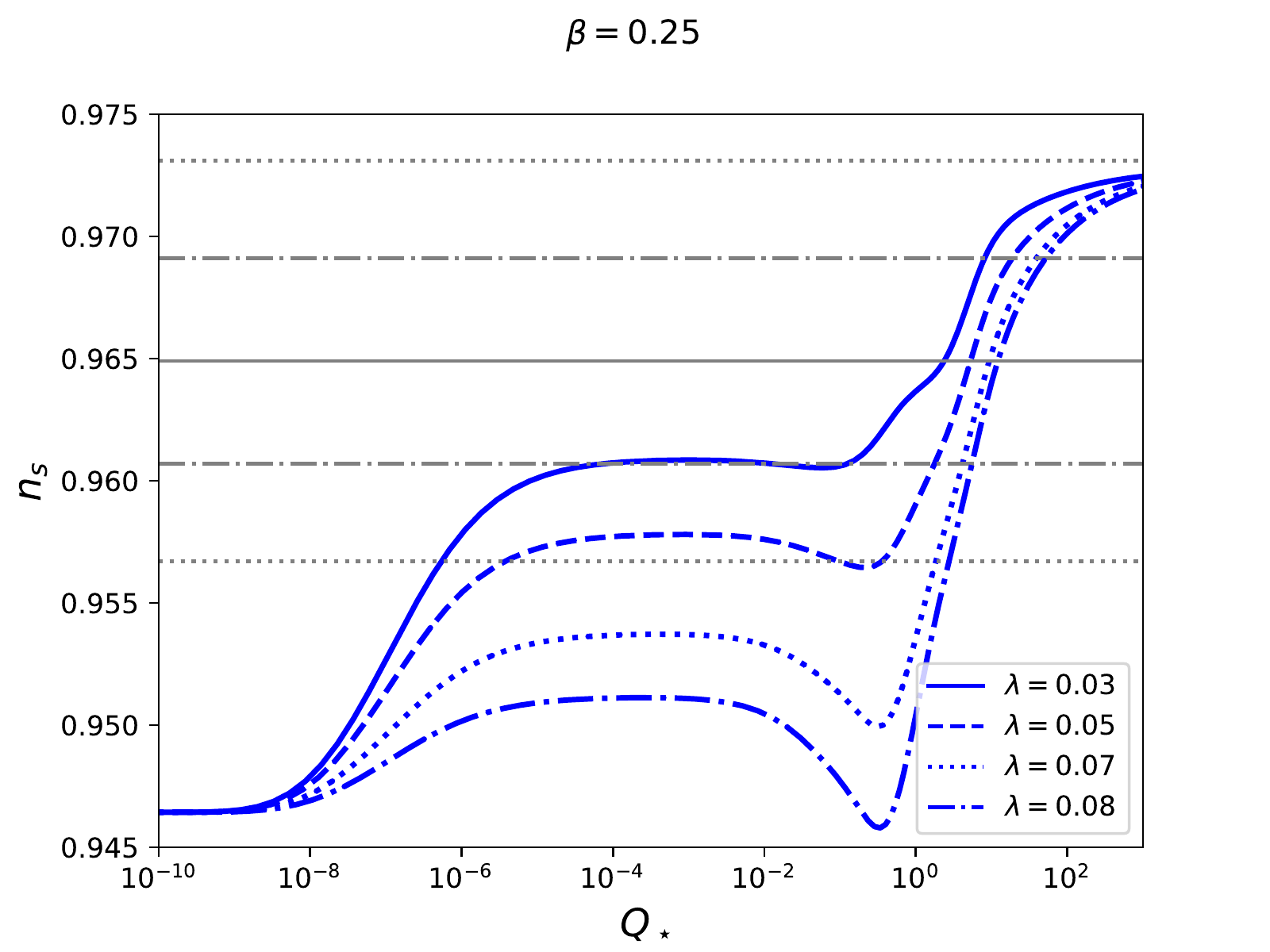}
	\includegraphics[width=7cm]{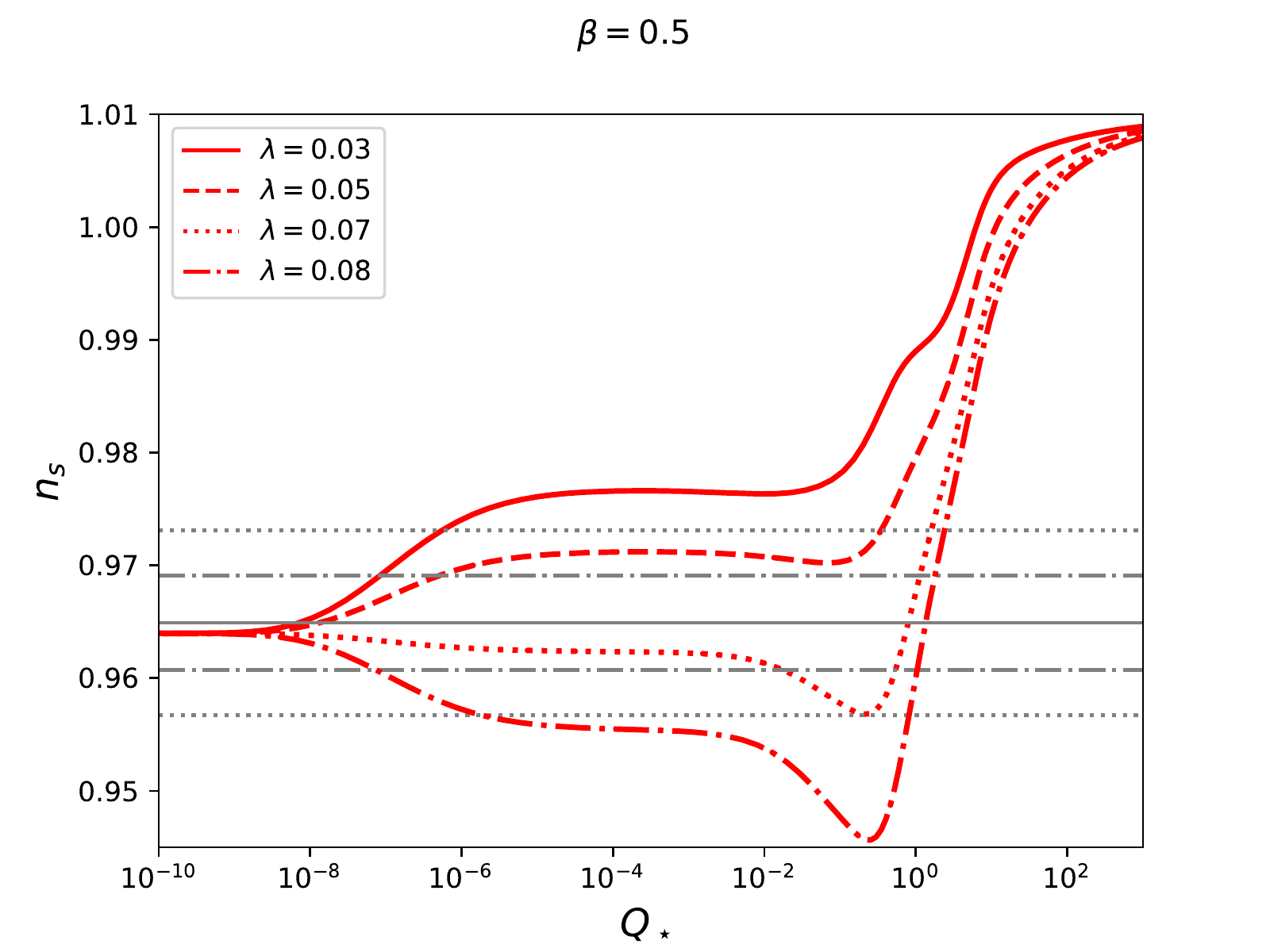}
	\includegraphics[width=7cm]{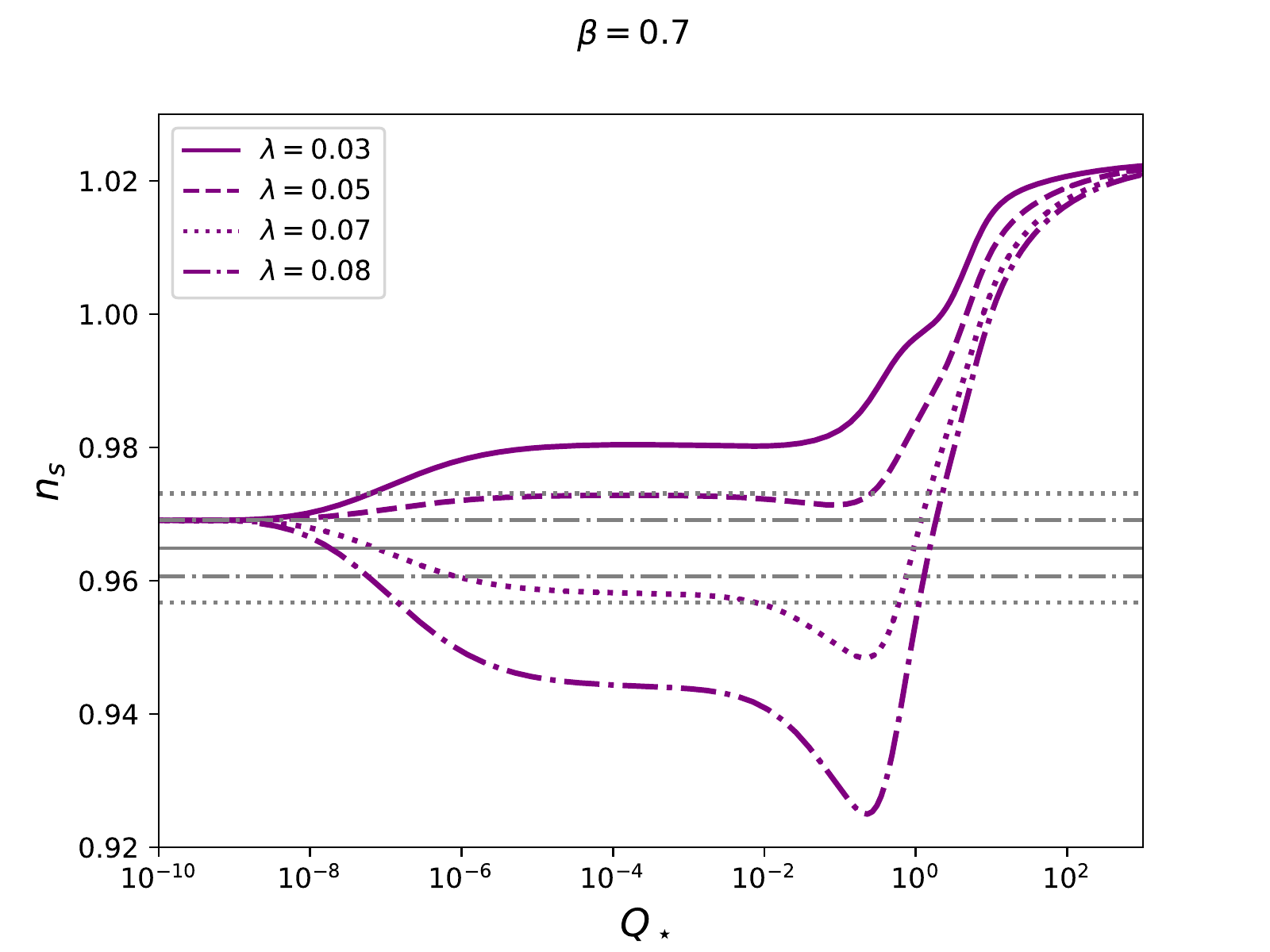}
	\caption{The spectral index $n_s$ as a function of $Q_\star$, for fixed values of $\beta$ and $\lambda=0.03,0.05,0.07,0.08$. As for the values of $\beta$, we have $\beta=0.2$ (upper left), $\beta=0.25$ (upper right), $\beta=0.5$ (lower left) and $\beta=0.7$ (lower right). The grey horizontal lines denote the 68\%,95\% limits of $n_s$ for Planck TT,TE,EE+lowl+lowE+lensing, $n_s=0.9649^{+0.0042+0.0082}_{-0.0042-0.0082}$ \cite{Planck:2018jri}. }
	\label{fig2}
\end{figure*}

Proceeding with the predictions of the $\beta$-exponential model, one can find that the relation between $Q$ and $\phi$ is given by

\begin{gather}
	Q(1+Q)^6=\mathcal{A}\lambda^6\left(\frac{M_p}{\phi}\right)^8\left(1-\lambda\beta\frac{\phi}{M_p}\right)^{1/\beta-6},
	\label{17}
\end{gather}
with
\begin{gather}
	\mathcal{A}\equiv\frac{C_\phi^4}{576\tilde g_\star^3}\left(\frac{V_0}{M_p^4}\right),
	\label{18}
\end{gather}
being another constant that encapsulates the dependence of $C_\phi$ with the amplitude $V_0$. The differential equation for the evolution of $Q$ in this case is
\begin{gather}
\frac{dQ}{dN} = \frac{\lambda^2Q}{(1+7Q)(1-\lambda\beta\phi/M_p)^2}\left[6\beta-1-\frac{\left(1-\lambda\beta\phi/M_p\right)M_p}{\lambda\phi}\right],
\label{19}
\end{gather}
meaning that while 
\begin{gather}
	\frac{\phi}{M_p} > \frac{1}{\lambda\left(7\beta-1\right)},
	\label{20}
\end{gather}
the growth of $Q$ with $N$ is ensured. We then use Eq. (\ref{17}) along with the slow-roll condition at the end of inflation $\epsilon_W=1$ to compute $\phi_{end}$ and $Q_{end}$ and evolve the differential equations for $\phi$, in Eq. (\ref{4}) and for $Q$, in Eq. (\ref{19}).

\begin{figure*}
	\includegraphics[width=8cm]{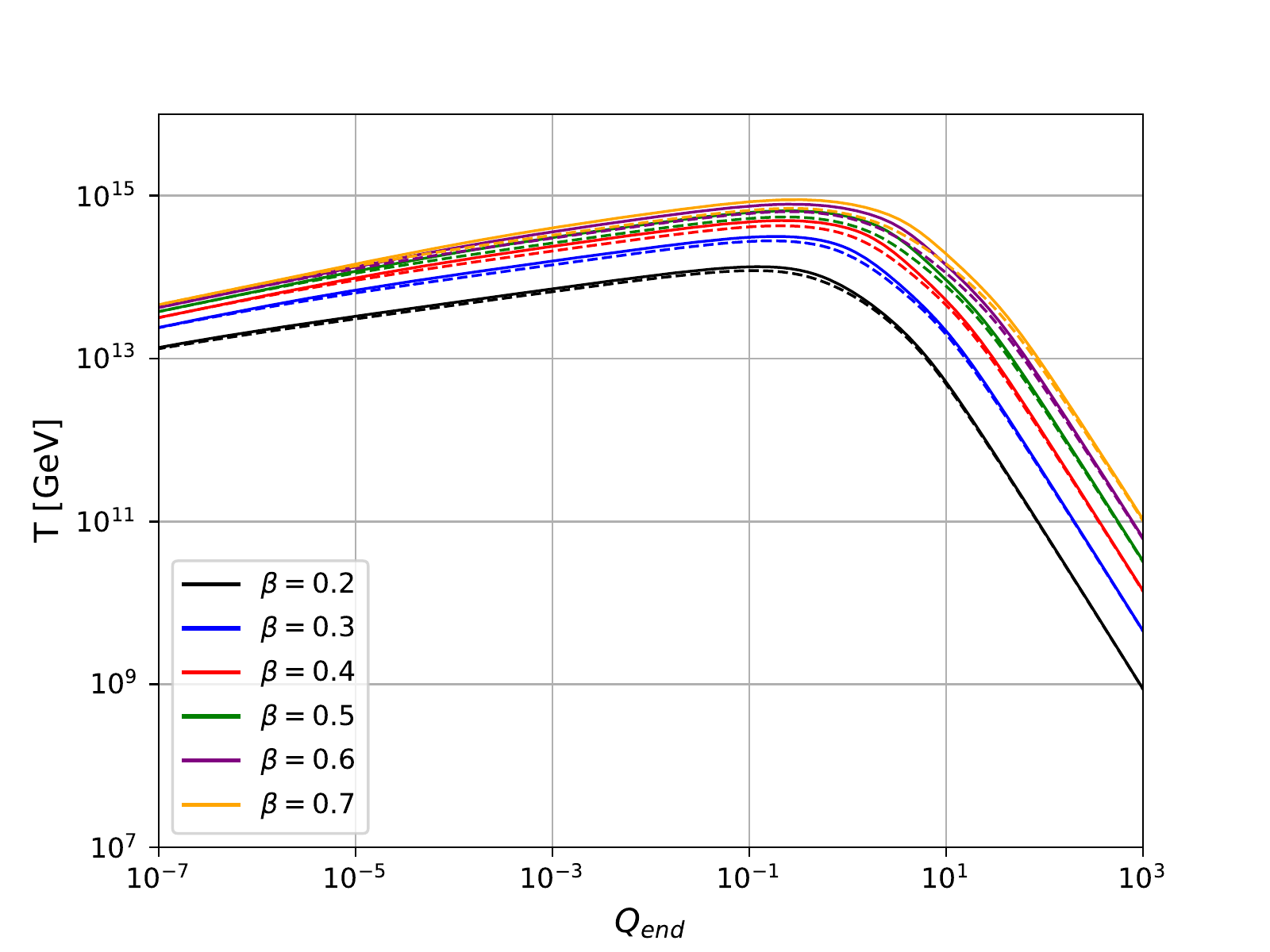}
	\includegraphics[width=8cm]{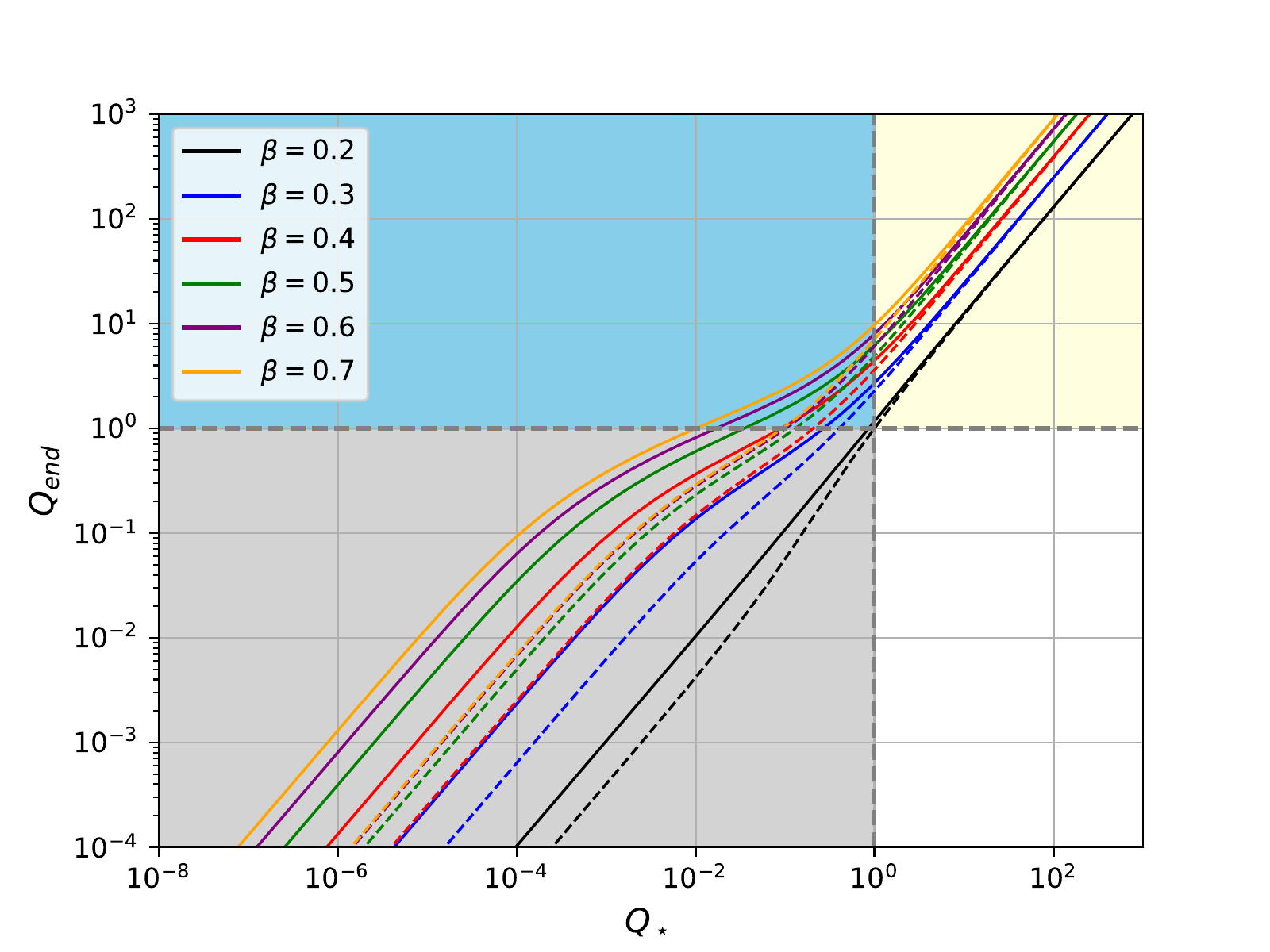}
	\caption{On the left panel, we plot the temperature of the thermal bath in GeV as a function of $Q_{end}$, for the same values of $\beta$ as in Fig. \ref{fig1}. On the right panel, we have the $Q_\star-Q_{end}$ plane, indicating inflation in the weak regime (grey region), strong regime (yellow region), and a transition from weak to the strong regime during inflation (blue region). Solid lines correspond to $\lambda=0.05$, while dashed lines correspond to $\lambda=0.07$.}
	\label{fig3}
\end{figure*}

The predictions of the $n_s$ and $r$ parameters are shown in Fig. \ref{fig1}. Choosing values of $\beta$ in the interval $0.2-0.7$, we have checked for any dependence on the $\lambda$ chosen; we see immediately that $\lambda$ greatly impacts the parameters. Considering the figure at the upper left, where $\lambda=0.05$, we note that larger values of $\beta$ are favored in the weak dissipative regime, as seen in the $\beta=0.3$ curve (blue dotted line). An effect of choosing higher values of $\lambda$ is shifting curves to lower $n_s$, favoring higher values of $\beta$ (dotted curves on the upper right panel), where we have chosen $\lambda=0.07$. It is then clear the importance of $\lambda$, as it can completely change the character of a curve for a given $\beta$. When we look at the strong dissipative regime, represented by the solid lines, a lower value of $\beta$ is preferred for both choices of $\lambda$. This happens because, as a higher $Q$ results in an increase of the spectral index for the dissipation coefficient we are considering, we need a value of $\beta$ that results in a low enough $n_s$ in the weak dissipation regime so that the curve will enter the confidence regions when $Q_\star>1$. From this, we can conclude that a small deviation from the exponential form is enough for inflation to end while also happening in a strong regime. Also, in Fig. \ref{fig1}, we plot the model's predictions for the running of the spectral index $n_{run}$ (lower panels). The effect of increasing $Q_\star$ is known \cite{Benetti:2016jhf}, in which the running can become positive for a larger $Q_\star$, but resulting in a spectral index too large, as seen in the figures. On the other hand, if $\beta$ (the deviation from the exponential potential) is small enough, a strong dissipation is allowed by data, and we can have values for the running that are well into the Planck constraints. However, we also note that while it is not possible to achieve a positive running for the smallest values of $\beta$ considered, a negative $n_{run}$ is estimated by the most recent Planck results of $n_{run}=-0.0045 \pm 0.0067$ (Planck TT,TE,EE+lowE+lensing). 

The difference between choices of $\lambda$ in the spectral index is more clearly shown in Fig. \ref{fig2}, where we plot the $n_s-Q_\star$ plane. Four values of $\beta$ are chosen, and we note how the curves change when we increase or decrease $\lambda$. When $\beta$ is low, it is challenging to achieve concordance with data in the weak dissipative regime, as the resulting $n_s$ is below the $2\sigma$ confidence limit, as seen in the $\beta=0.2$ plot (upper left panel). As we go towards the strong dissipative regime, however, it is possible for the curves to reach the confidence region since there is an increase of $n_s$. By slightly increasing $\beta$, we can accommodate both regimes into the constraints for $n_s$, depending on the $\lambda$ chosen (upper right panel). It results in an upward shift in curves for $\beta=0.25$ choice, in which the shape of curves for each $\lambda$ is essentially the same. This changes, however, for higher values of $\beta$. Choosing $\beta=0.5$ (lower left panel), it is clear how the choice of $\lambda$ affects the curves compared to the two previous choices, and the same goes for $\beta=0.7$ (lower right panel). The main results for all these choices are the increase of $n_s$ to a higher constant value at $Q_\star\gg 1$, and the convergence of all lines for a very low/high $Q_\star$, indicating that $\lambda$ has a negligible effect for more extreme values. Finally, we note that for $\beta\geq 1/2$, it is in general difficult to achieve concordance with data in the strong dissipative regime for $N_\star=55$, which is in conflict with the restriction for $\beta$ given in \cite{Santos:2017alg}; on the other hand, there is an easy agreement with the Planck limits, when we consider inflation taking place in the weak regime.

In Fig. \ref{fig3}, we plot the dependence of the temperature at the end of inflation with $Q_{end}$ (left panel) and the $Q_\star-Q_{end}$ plane (right panel). Looking at $T$, we first note that while the dependence on $\beta$ is more evident, the impact of $\lambda$ in the curves is not as strong as it is in the $n_s-r$ plane: we have almost the same predictions for both values of $\lambda$, indicated by the solid and dotted curves.  The general behavior is as follows: if inflation ends in the weak regime, the temperature is high, varying from $T\sim 10^{13}-10^{15}$ GeV, and an increase with $Q_{end}$ is noticeable. However, as soon as $Q_{end}\sim 1$, the picture changes completely. In the interval $Q_{end}\sim 1-1000$, the temperature decreases by many order of magnitude, being able to reach $T\sim 10^9$ GeV for $\beta=0.2$ and $Q_{end}\sim 10^3$. As for the impact of $\beta$, we see that higher $\beta$ will correspond to higher $T$, but there seems to be a maximum value for the temperature for each $Q_{end}$ as the curves become closer to one another as $\beta$ increases. Looking now at the $Q_\star-Q_{end}$ plane in the right panel, it is possible to see the relation between the values of $Q$ with inflation proceeding totally in the weak/strong regime or when a transition between regimes happens. Curves inside the grey region correspond to a state where inflation happens entirely in the weak dissipation regime, while the yellow region corresponds to inflation in the strong regime. Curves represent a transition from weak to strong regimes are in the blue region, where we note that $Q_\star$ must be at least of order $10^{-2}$, and the higher $\beta$ increase the parameter space for which the transition can happen. 

\begin{figure*}
	\centering
	\includegraphics[width=8cm]{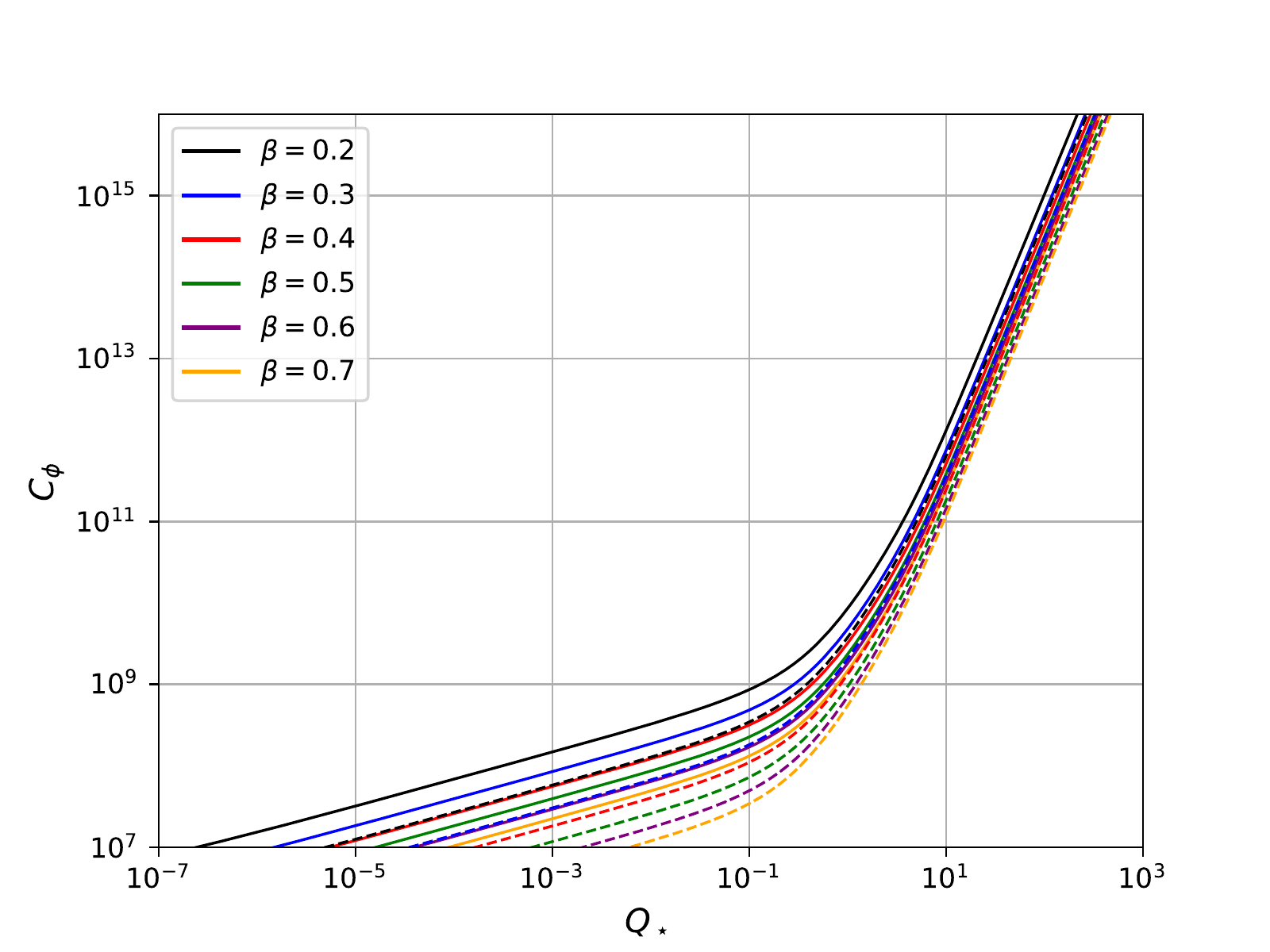}
	\includegraphics[width=8cm]{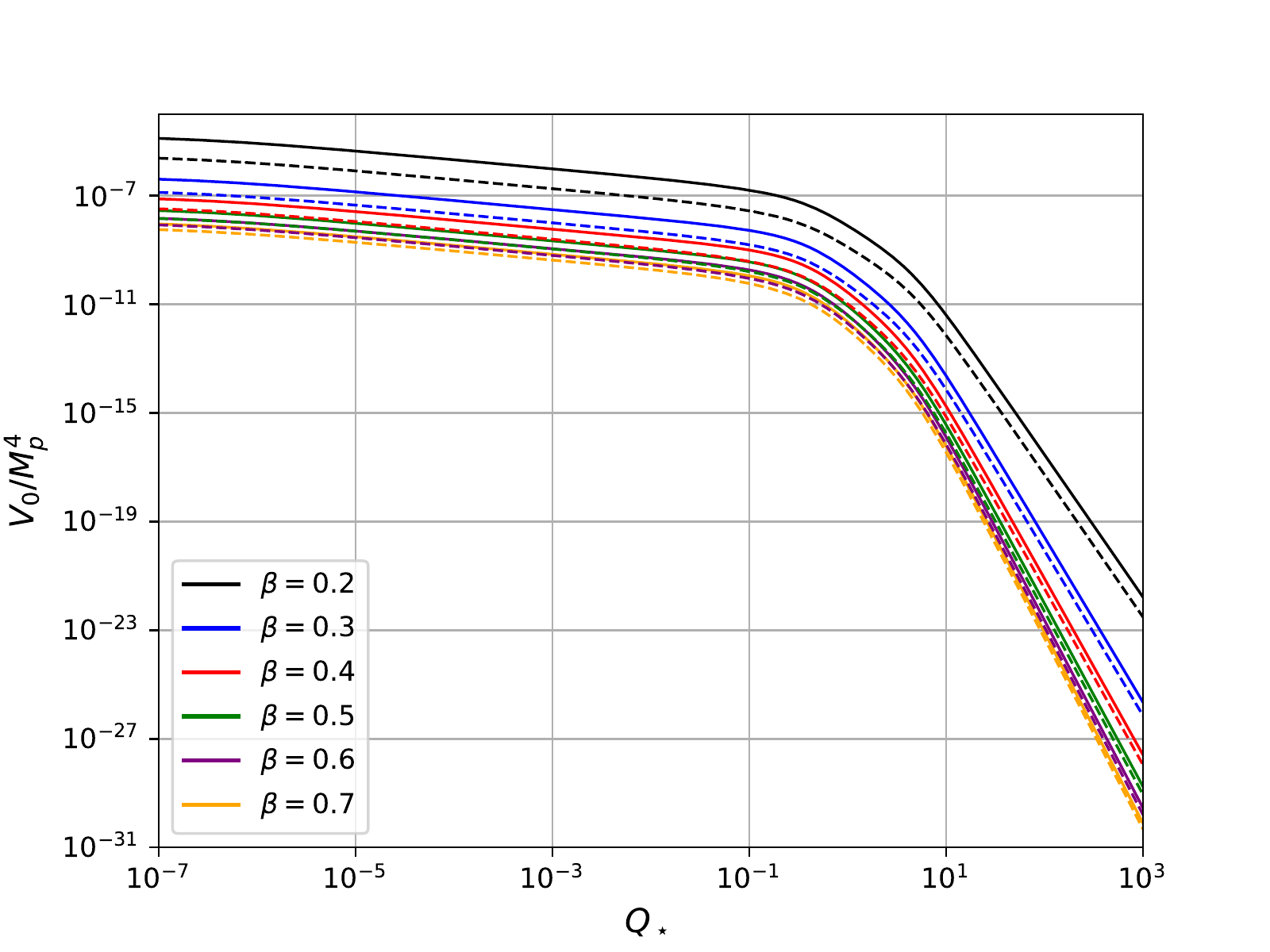}
	\caption{The coefficient $C_\phi$ is shown as a function of $Q_\star$, for the same values of $\beta$ as in fig. \ref{fig1} (left panel). On the right panel, we have the amplitude of the potential $V_0$ also as a function of $Q_\star$. Once more, solid lines correspond to $\lambda=0.05$, while dashed lines correspond to $\lambda=0.07$.}
	\label{fig4}
\end{figure*}

Finally, we show in Fig. \ref{fig4} how the coefficient $C_\phi$ (left panel) and the potential amplitude $V_0$ (left panel) vary for a wide range of $Q_\star$. We remember that $C_\phi$ can be estimated from Eq. (\ref{18}), while $V_0$ is obtained by fixing the amplitude of scalar perturbations in Eq. (\ref{8}). In a supersymmetric implementation, $C_\phi$ is interpreted as directly proportional to the product of multiplicities of chiral fields $X$ and $Y$ present in the model \cite{Bastero-Gil:2012akf}. We note that a huge multiplicity is present in warm inflation models, where in particular, the number increases sharply in the strong dissipation regime, of order $N_XN_Y\sim 10^{15}$. This result was pointed out in \cite{Das:2020xmh}, which is also confirmed here, and this can be seen as a problem associated with the warm inflation scenario. The different behavior between values of $Q_\star$ is also noticed when one estimates the amplitude of the potential $V_0$, shown in the left panel of Fig. \ref{fig4}. A decrease of several orders of magnitude is seen when inflation occurs in the strong regime, ranging from $V_0\sim 10^{-23}$ to $V_0\sim 10^{-31}$ for the values of $\beta$ chosen.

\section{Is this model in accordance with the swampland conjectures?}\label{sec4}

Since the establishment of quantum gravity theories, such as string theory, efforts have been made towards
finding scenarios consistent within this context, including cosmological models. These continuous attempts ended up giving rise to what is known today as the \textit{swampland conjectures}, which are translated into conditions that can establish bounds in many models, including scalar field theory-based ones. The first conjecture was determined from the search for a stable de-Sitter vacuum in string theory  \cite{Danielsson:2018ztv,Dine:2020vmr,Kachru:2003aw} resulting in the \textit{de Sitter} (dS) Swampland criterion. This discussion is important in cosmology since we know two periods of accelerated universe expansion: inflation and dark energy domination. Both lead the universe to a state close to a de Sitter one. As more developments were made, other swampland criteria were proposed. The swampland \textit{distance} conjecture \cite{Ooguri:2006in} gives an upper limit on the excursion of the scalar field along the potential during inflation. These limits are important in the context of a universe whose acceleration is described by the evolution of a scalar field, which is (possibly) the case of inflation, and possible candidates for late-time acceleration, such as quintessence models. A third condition also has appeared, which restricts sub-Planckian modes from leaving the horizon during inflation, the \textit{Transplanckian Censorship} conjecture (TCC) \cite{Bedroya:2019snp,Bedroya:2019tba}. These conjectures have been used to restrict inflationary models in the past years. However, it was seen that the usual picture of canonical, single-field inflation that leaves a cold universe afterwards is in direct conflict with all of the conjectures. In this way, if one wants to conciliate slow-roll inflation with more general theories, it might be necessary to consider significant changes in how we view the early universe's evolution.

It is possible to summarize the three criteria as conditions imposed on an inflationary potential. They are expressed as

\begin{itemize}
	\item dS conjecture: this criterion limits the slope of a given scalar potential in a way that the gradient of $V$ must follow
	\begin{gather}
	   M_{p}\left(\frac{\vert\nabla_\phi V\vert}{V}\right)>a
	   \label{21}
	\end{gather}
	with $a\simeq\mathcal{O}(1)$. We see here the first problem: during inflation, both slow-roll parameters must be much less than one, so the simplest picture of inflation cannot satisfy this conjecture regardless of the model. This condition favors steep potentials, which are usually unsuitable for inflation; at the same time, it excludes potentials with plateaus, which are the most favored by data, such as the Starobinsky model \cite{Motaharfar:2018zyb}. 
	
	\item Distance conjecture: this conjecture gives a restriction on the excursion of the scalar field during inflation; essentially
	\begin{gather}
		\frac{\vert\Delta\phi\vert}{M_p}<b,
		\label{22}
	\end{gather}
	with $b$ being another constant of order one. It restricts classes of inflationary models in which inflation happens. The field excursion is not super-Planckian; in particular, this conjecture tells that small field models of inflation might be favored, while large field models, such as those given by monomial potentials, are disfavored.
	
	\item TCC: by placing a limit on the scales that leave the horizon during inflation, one gets a limit on the energy scale of inflation itself
	\begin{gather}
		V^{1/4} < 3\times 10^{-10}M_p.
		\label{23}
	\end{gather}
	This bound comes from the imposition that scales of the order of Planck length should not leave the horizon during inflation. As a result, such limitation results in a tensor-to-scalar ratio of order $r\sim 10^{-30}$, which might be a problem from the observational standpoint, while stating that the primordial gravitational wave spectrum is significantly weaker when compared to the one produced by scalar perturbations.	
\end{itemize}

\begin{figure*}
	\centering
	\includegraphics[width=8cm]{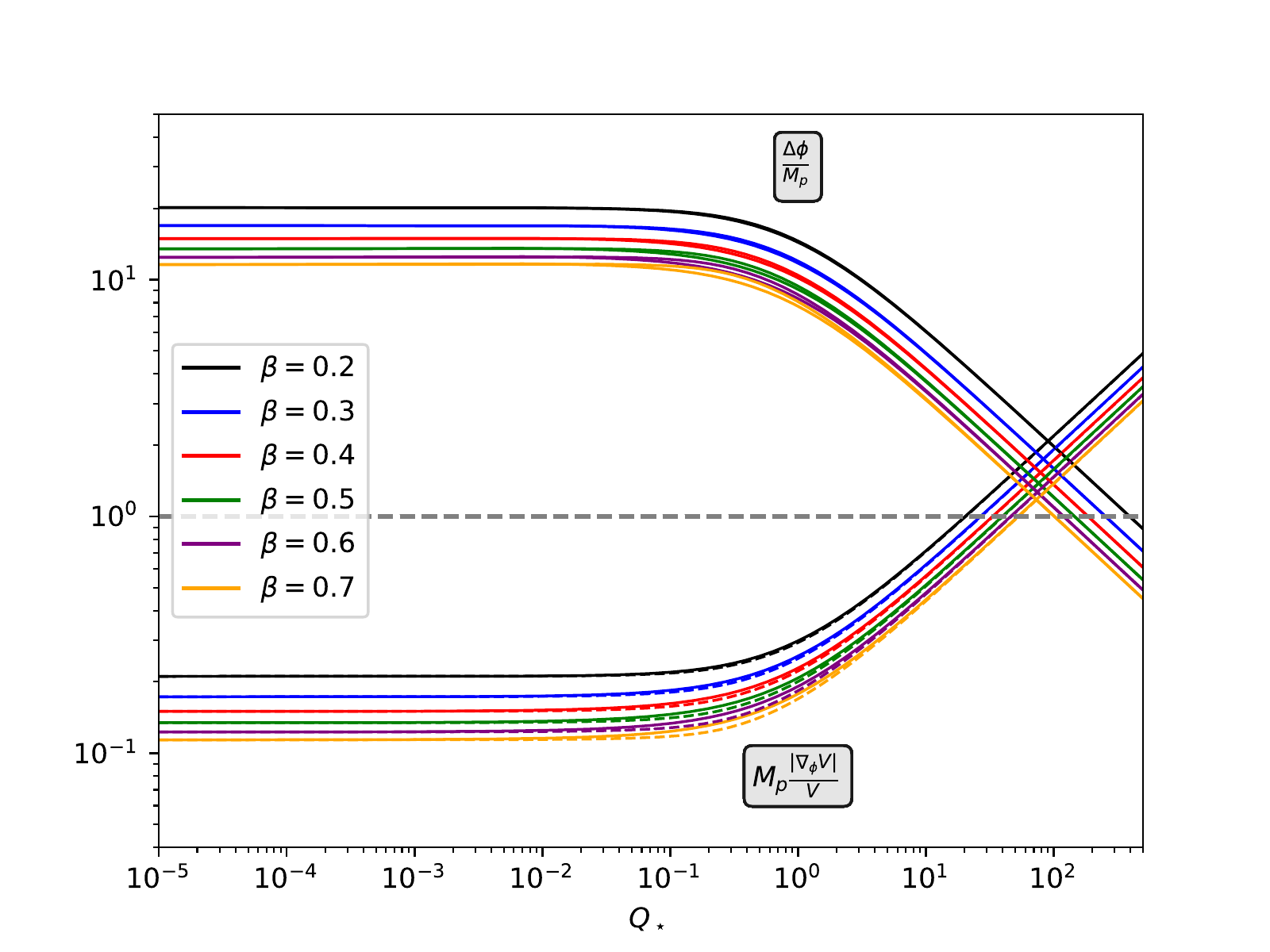}
	\includegraphics[width=8cm]{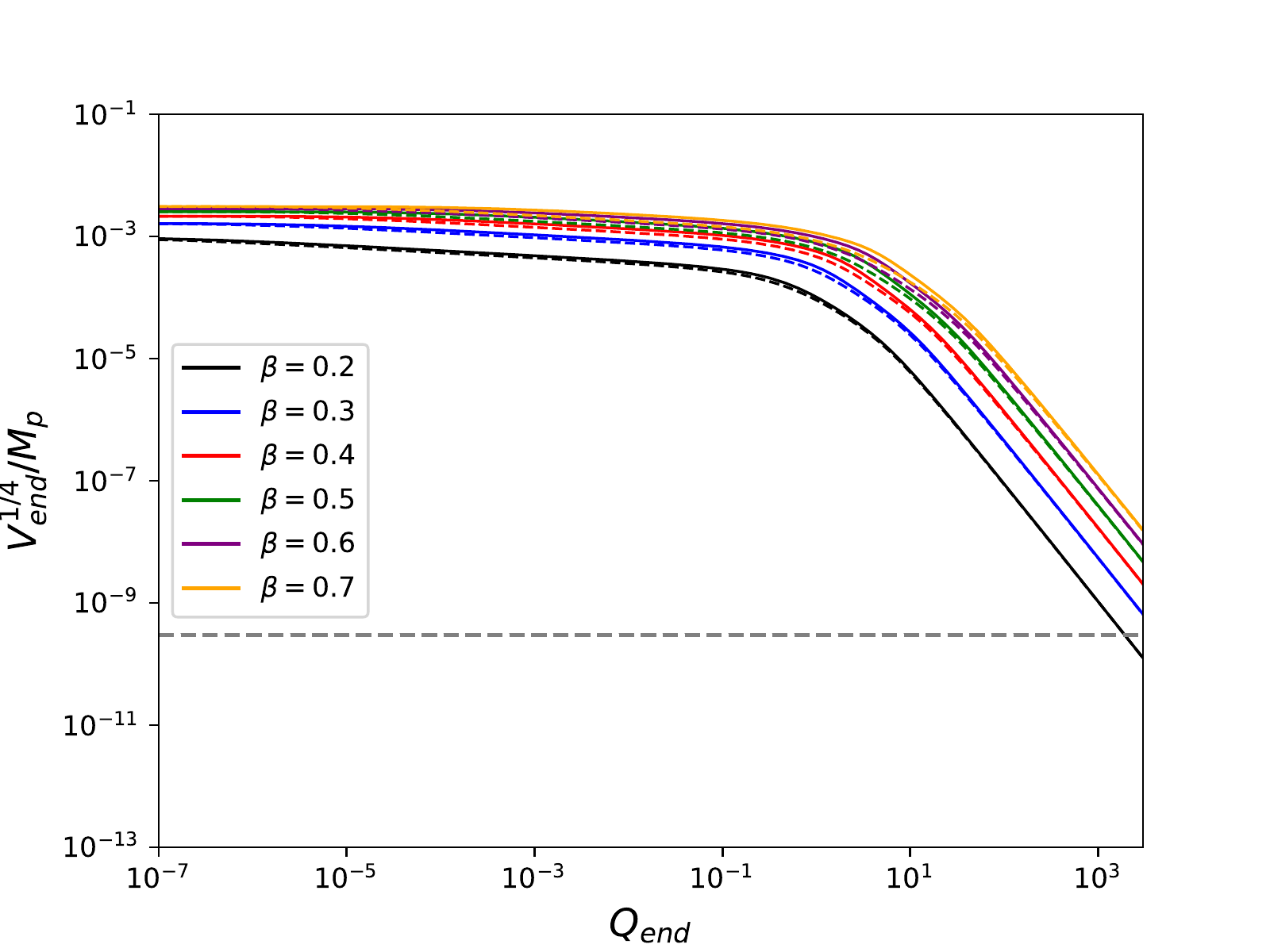}
	\caption{The ratio $M_p|V_{,\phi}|/V$, the field excursion $\Delta\phi$ (left panel) and the potential $V_{end}^{1/4}/M_p$ (right panel) as a function of $Q_\star$ for the $\beta$-exponential model when the dissipation coefficient is given by Eq. (\ref{16}). We have fixed $\lambda=0.05$ (solid lines), $\lambda=0.07$ (dashed lines) and have considered different $\beta$ in the range $\beta=0.2-0.7$.}
	\label{fig5}
\end{figure*}

Recent works in the literature attempted to fulfill the three conditions in extensions of the usual inflationary scenario. In particular, the idea of warm inflation comes as a possibility to realize inflation while satisfying the swampland conditions. For instance, by looking at (\ref{5}), we see that due to the $(1+Q)$ factor, in the strong dissipative regime, it is possible to have $\epsilon_V,\eta_V>1$, as imposed by the dS conjecture, at the same time that $\epsilon_W,|\eta_W|\ll 1$, which are the conditions for slow-roll inflation. Also, the presence of the $\Gamma\dot\phi$ term in Eq. (\ref{1}) means that extra friction is being added as the field rolls down the potential; as a consequence, it can be possible for a larger class of models to be consistent with the distance conjecture (Eq. \ref{22}).

Fig. \ref{fig5} shows how the warm $\beta$-exponential model behaves when confronted with the swampland conjectures. We have considered the same range of $\beta$ as described in the $n_s-r$ plane while fixing $\lambda=0.05,0.07$, as we have found that the results are very similar when other values are considered. We first plot $|V_\phi/V|$ in the left panel. While in the weak dissipation regime ($Q_\star<1$), the smallness of the quantity is guaranteed, as soon as $Q_\star\approx1$, it starts rising to values that can be larger than one. We note that smaller $\beta$ leads to $|V_\epsilon/V|=1$ being reached for a smaller $Q_\star$. In the strong regime where $Q_\star\geq10$, the dS conjecture is easily satisfied for all values of $\beta$ considered, so we have the freedom to choose some $\beta$ that are in concordance with the Planck constraints on $n_s$ and $r$. Next, we check if the model satisfies the distance conjecture by showing the field excursion $\Delta\phi/M_p$ as a function of $Q_\star$. In the weak dissipation, the excursion is super-Planckian, but again, when $Q_\star$ approaches unity, $\Delta\phi/M_p$ decreases, so that when $Q_\star\geq 10$, the field excursion becomes sub-Planckian, as the dissipation is large enough to make even a steep potential to behave like a small-field model. As for the last conjecture, TCC is showed on the right panel of Fig.\ref{fig5}, where we can see that $V_{end}^{1/4}/M_p$ is almost constant in the weak dissipative regime, but decreases sharply in the strong regime by many orders of magnitude, especially for the cubic coefficient case; here, the TCC is respected for $Q_\star\geq10^3$.

\section{Discussion and Conclusions}\label{sec5}

The warm inflation picture has attracted much attention in the past years as the reheating process is surrounded by many questions on its realization. This is mainly due to the fact that it is a difficult epoch to probe and construct consistent connections with the inflation era from a microscopical perspective. Also, well-motivated inflationary models, such as those given by potentials of monomial and exponential form, suffer from inconsistency with CMB data. Indeed, when one looks to inflationary parameters $n_s$ and $r$, some extension that allows concordance with data to be restored is usually required. While for instance, a non-minimal coupling of the inflaton with gravity might alleviate much of these issues \cite{dosSantos:2021vis,Campista:2017ovq}, it is interesting (and plausible) to consider a situation where the energy stored in the inflaton is converted to other particles as inflation happens, resulting in a warm inflationary universe that can lead towards a radiation-dominated universe afterwards. Another issue related to inflationary models is their agreement with the recently proposed swampland conjectures \cite{Das:2018hqy,Das:2019acf}, which result from attempts to incorporate models based on field theory into general ones such as string theory. In general, minimally-coupled scalar field inflationary models are inconsistent with these conjectures due to the essential smallness of the slow-roll parameters for inflation or a large energy scale of inflation. Recent works \cite{Das:2019acf,Das:2020xmh,Kamali:2019xnt} have looked into the conjectures from a warm inflation perspective. It was found that the three proposed conjectures can be simultaneously satisfied when inflation takes place in a strong dissipation regime. At the same time, cosmological parameters such as the spectral index and tensor-to-scalar ratio can be driven into the Planck constraints. 

In this work, we have considered how the $\beta$-exponential inflationary model \cite{Alcaniz:2006nu,Santos:2017alg,dosSantos:2021vis} behaves when we allow dissipation of the inflaton energy into radiation in the warm inflation picture. In the original model, for $N_\star=50-60$, it is not possible to have a low enough $r$ while satisfying the constraints for $n_s$ \cite{dosSantos:2021vis}. In contrast, in the warm inflation scenario, we can easily have concordance with data when inflation happens in the weak dissipation regime for a restricted interval of $\beta$. We have considered a dissipation coefficient, in which a cubic dependence on the temperature exists. It is found that both weak and strong dissipation regimes are allowed, and in particular, the strong regime leads $n_s$ to a constant value as $Q_\star$ increases, with little dependency on the $\lambda$ parameter. This case is then viewed favorably in light of the swampland conjectures, as we have found (see Fig. \ref{fig5}) that all three conjectures are satisfied in the strong regime for the values of $\beta$ considered.

This topic is far from over and further investigations are underway. In fact, considering a recently proposed warm inflation scenario where brane effects are considered for an exponential potential \cite{Kamali:2019xnt}, in which inflation ends more naturally, the swampland conjectures are satisfied while also providing a connection with the late-time universe as a quintessence model. We consider it interesting to see how a deviation from the exponential form might affect the results. Another interesting  and important aspect of our investigation is the study of how these models, especially in the strong dissipation regime, affect the predictions for the CMB power spectra.  All these questions are being investigated and will be discussed in an upcoming paper.

\section*{Acknowledgements}

We thank Prof. Rudnei Ramos for fruitful discussions. 
F.B.M. dos Santos is supported by Coordena\c{c}\~{a}o de Aperfei\c{c}oamento de Pessoal de N\'ivel Superior (CAPES). R. Silva acknowledges financial support from CNPq (Grant No. 307620/2019-0). SSC is supported by the Istituto Nazionale di Fisica Nucleare (INFN), sezione di Pisa, iniziativa specifica TASP. MB is supported by the INFN, sezione di Napoli, iniziativa specifica QGSKY. J. Alcaniz is supported by Conselho Nacional de Desenvolvimento Cient\'{\i}fico e Tecnol\'ogico CNPq (Grants no. 310790/2014-0 and 400471/2014-0) and Funda\c{c}\~ao de Amparo \`a Pesquisa do Estado do Rio de Janeiro FAPERJ (grant no. 233906).

\appendix

\section{Deriving $n_s$ for a given $\Gamma$}\label{appA}

To obtain the spectral index $n_s$ by following Eq. (\ref{12}), we can first write Eq. (\ref{8}) in the form
\begin{gather}
	\operatorname{log}\Delta^2_\mathcal{R} = \operatorname{log}P_\mathcal{R} + \operatorname{log}\left(1+2n_{BE}+f(Q)\frac{T}{H}\right) + \operatorname{log}G(Q),
	\label{a1}
\end{gather}
where
\begin{gather}
	P_\mathcal{R} = \frac{H^2}{4\pi^2\phi^{'2}}, \quad f(Q) \equiv \frac{2\sqrt{3}\pi Q}{\sqrt{3+4\pi Q}},
	\label{a2}
\end{gather}
and differentiate everything with respect to $N$, denoted by the primes. For the first term of (\ref{a1}), with the help of eq. (\ref{2.4}), we find that
\begin{align}
	\frac{d\operatorname{log}P_\mathcal{R}}{dN} &= 2\frac{d\operatorname{log}H}{dN} - 2\frac{d\operatorname{log}\phi'}{dN},\nonumber\\
	&= -6\epsilon_w+ 2\eta_w + \frac{2Q'}{1+Q},
	\label{a3}
\end{align}
For the second term, we note that
\begin{gather}
	\frac{d}{dN}\left(1+2n_{BE}+f(Q)\frac{T}{H}\right)=\left[2n'_{BE}+\frac{T}{H}f'+f\left(\frac{T}{H}\right)'\right],
	\label{a4}
\end{gather}
in which 
\begin{align}
	n'_{BE} =& \frac{2e^{H/T}(T/H)'}{(e^{H/T}-1)^2(T/H)^2},
	\label{a5}
\end{align}	
\begin{align}	
	 \frac{T}{H}f'=&\left(\frac{T}{H}\right)\frac{df}{dQ}Q' \nonumber\\  =&\left(\frac{T}{H}\right)\frac{2\sqrt{3}\pi(2\pi Q+3)}{(3+4\pi Q)^{3/2}}Q'.
	 \label{a6}
\end{align}
Finally, from the last term of (\ref{a1}) we have simply
\begin{gather}
	\frac{d\operatorname{log}G}{dN} = \frac{1}{G}\frac{dG}{dQ}Q'.
	\label{a7}
\end{gather}
Gathering all the terms together, we find that
\begin{align}
	n_s-1 & = -6\epsilon_w + 2\eta_w + \frac{2Q'}{1+Q} +  \frac{1}{G}\frac{dG}{dQ}Q'
	\nonumber\\ & + \frac{1}{\left(1+2n_{BE}+f(Q)\frac{T}{H}\right)}\Bigg[\frac{2e^{H/T}(T/H)'}{(e^{H/T}-1)^2(T/H)^2}   \nonumber\\ & + \frac{T}{H}\frac{2\sqrt{3}\pi(2\pi Q+3)}{(3+4\pi Q)^{3/2}}Q' + \frac{2\sqrt{3}\pi Q}{\sqrt{3+4\pi Q}}\left(\frac{T}{H}\right)'\Bigg].
	\label{a8}
\end{align}

To compute $n_s$ for a specific $\Gamma$, we use the evolution equations \cite{Bastero-Gil:2009sdq,Bastero-Gil:2016qru} for $\phi$ and $Q$ that must be substituted in (\ref{a8}):

\begin{gather}
	Q' = \frac{Q}{1+7Q}\left(10\epsilon_V-6\eta_V+8\sigma_V\right),
	\label{a9}
\end{gather}
\begin{gather}
	\left(\frac{T}{H}\right)' = \frac{2(T/H)}{1+7Q}\left(\frac{2+4Q}{1+Q}\epsilon_V - \eta_V + \frac{1-Q}{1+Q}\sigma_V\right),
	\label{a10}
\end{gather}
with $\sigma_V\equiv M_p^2 V_{,\phi}/(\phi V)$.

\section{Deriving $n_{run}$ for a given $\Gamma$}\label{appB}

As the running involves second derivatives, it is useful to determine first the differential equations for $Q''$,$\epsilon_V'$, $\eta_V'$, $\sigma_V'$ and $(T/H)''$:
\begin{gather}
Q'' = \frac{Q(10\epsilon'_V - 6\eta'_V + 8\sigma'_V)}{1+7Q} + \frac{Q^{'2}}{Q} - \frac{7Q^{'2}}{1+7Q},
\end{gather}

\begin{gather}
 \epsilon'_V = \frac{2\epsilon_V}{1+Q}\left( 2\epsilon_V - \eta_V \right),   
\end{gather}

\begin{gather}
 \eta'_V = \frac{1}{1+Q}\left(2\epsilon_V\eta_V - \zeta^2 \right), \quad \zeta^2\equiv M_p^2\frac{V_{,\phi\phi\phi}V_{,\phi}}{V^2},
\end{gather}

\begin{gather}
\sigma'_V = \frac{1}{1+Q}\left( 2\epsilon_V\sigma_V - \eta_V\sigma + \sigma_V^2  \right)
\end{gather}

\begin{align}
\left(\frac{T}{H}\right)'' & = 2\left[ \frac{(T/H)'}{1+7Q} - \frac{7Q'(T/H)}{(1+7Q)^2} \right]\left[ \frac{(2+4Q)\epsilon_V}{1+Q} - \eta_V + \frac{1-Q}{1+Q}\sigma_V \right] &\nonumber\\& + 2\frac{(T/H)}{1+7Q}\Big[ \frac{2+4Q}{1+Q}\epsilon'_V + \frac{4\epsilon_V}{1+Q}Q' - \frac{(2+4Q)\epsilon_V}{(1+Q)^2}Q' - \eta'_V \nonumber\\& - \frac{\sigma_V}{1+Q}Q' + \frac{1-Q}{1+Q}\sigma'_V - \frac{(1-Q)\sigma_V}{(1+Q)^2}Q' \Big].
\end{align}

As we approximate the running as $n_{run}\simeq\frac{d^2\log\Delta^2_\mathcal{R}}{dN^2}$, from \ref{a1}, we compute 
\begin{gather}
n_{run} = \frac{d^2\log P_\mathcal{R}}{dN^2} + \frac{d^2\log F}{dN^2} + \frac{d^2\log G}{dN^2}, 
\end{gather}	
\begin{gather}		
F\equiv 1+2n_{BE}+f(Q)\frac{T}{H},
\end{gather}
where the individual terms can be obtained as
\begin{gather}
\frac{d^2\log P_\mathcal{R}}{dN^2} =  \frac{(6\epsilon_V - 2\eta_V - 2Q')}{(1+Q)}Q' + \frac{2\eta'_V - 6\epsilon'_V + 2Q''}{1+Q}
\end{gather}

\begin{gather}
\frac{d^2\log G}{dN^2} = \frac{Q^{'2}}{G}\frac{d^2G}{dQ^2} + \frac{dG}{dQ}\frac{Q''}{G} - \frac{(Q'\frac{dG}{dQ})^2}{G^2}
\end{gather}

\begin{gather}
\frac{d^2\log F}{dN^2} = \frac{F''}{F} - \left(\frac{F'}{F}\right)^2,
\end{gather}
\begin{gather}	
F'\equiv 2n'_{BE} + \left(\frac{T}{H}\right)f'(Q) - f(Q)\left(\frac{T}{H}\right)',
\end{gather}
\begin{gather}	
F''\equiv 2n''_{BE} + \left(\frac{T}{H}\right)'f'(Q) + \left(\frac{T}{H}\right) f''(Q) + 
\end{gather}
\begin{gather}	
f'(Q)\left(\frac{T}{H}\right)' f(Q)\left(\frac{T}{H}\right)'',
\end{gather}
\begin{gather}	
f'(Q) = \frac{df}{dQ}Q',
\end{gather}
\begin{gather}	
f''(Q) = Q'\left[\frac{d^2f}{dQ^2}Q' + \frac{df}{dQ}\frac{Q''}{Q'}\right].
\end{gather}

\bibliography{references}

\end{document}